\documentclass[reprint,superscriptaddress,amssymb,amsmath,aps,prx,longbibliography]{revtex4-1}

\pdfoutput=1 
\usepackage{graphicx}
\usepackage{amssymb}
\usepackage{epstopdf}
\usepackage{upgreek}
\usepackage{dcolumn}
\usepackage{bm}
\usepackage{gensymb}
\usepackage{braket}
\usepackage{amsmath}
\usepackage{xcolor}
\usepackage{float}

\usepackage{tikz}

\expandafter\ifx\csname package@font\endcsname\relax\else
\expandafter\expandafter
\expandafter\usepackage
\expandafter\expandafter
\expandafter{\csname package@font\endcsname}%
\fi

\newcommand{\micro}{$\upmu$}

\newcommand{\micron}{$\upmu$m }

\newcommand{\be}{\begin{eqnarray}}
\newcommand{\ee}{\end{eqnarray}}
\newcommand{\bfig}{\begin{figure}}
	\newcommand{\efig}{\end{figure}}

\DeclareFontFamily{U}{mathb}{}
\DeclareFontShape{U}{mathb}{m}{n}{
	<-5.5> mathb5
	<5.5-6.5> mathb6
	<6.5-7.5> mathb7
	<7.5-8.5> mathb8
	<8.5-9.5> mathb9
	<9.5-11.5> mathb10
	<11.5-> mathbb12
}{}

\usepackage[colorlinks=true,allcolors=blue]{hyperref}%

\begin{document}
	\title{Resolving Phonon Fock States in a Multimode Cavity with a Double-Slit Qubit}
	\author{L. R. Sletten}
	\email{lucas.sletten@colorado.edu}
	\affiliation{JILA, National Institute of Standards and Technology and the University of Colorado, Boulder, Colorado 80309, USA}
	\affiliation{Department of Physics, University of Colorado, Boulder, Colorado 80309, USA}
	
	\author{B. A. Moores}
	\affiliation{JILA, National Institute of Standards and Technology and the University of Colorado, Boulder, Colorado 80309, USA}
	\affiliation{Department of Physics, University of Colorado, Boulder, Colorado 80309, USA}
	
	\author{J. J. Viennot}
	\affiliation{JILA, National Institute of Standards and Technology and the University of Colorado, Boulder, Colorado 80309, USA}
	\affiliation{Department of Physics, University of Colorado, Boulder, Colorado 80309, USA}

	\author{K. W. Lehnert}
	\affiliation{JILA, National Institute of Standards and Technology and the University of Colorado, Boulder, Colorado 80309, USA}
	\affiliation{Department of Physics, University of Colorado, Boulder, Colorado 80309, USA}
	\date{\today}

	\begin{abstract}		
We resolve phonon number states in the spectrum of a superconducting qubit coupled to a multimode acoustic cavity. Crucial to this resolution is the sharp frequency dependence in the qubit-phonon interaction engineered by coupling the qubit to surface acoustic waves in two locations separated by $\sim40$ acoustic wavelengths. In analogy to double-slit diffraction, the resulting interference generates high-contrast frequency structure in the qubit-phonon interaction. We observe this frequency structure both in the coupling rate to multiple cavity modes and in the qubit spontaneous emission rate into unconfined modes. We use this sharp frequency structure to resolve single phonons by tuning the qubit to a frequency of destructive interference where all acoustic interactions are dispersive. By exciting several detuned yet strongly coupled phononic modes and measuring the resulting qubit spectrum, we observe that, for two modes, the device enters the strong dispersive regime where single phonons are spectrally resolved. 
	\end{abstract}
	\maketitle

	
Quantum control over mechanical degrees of freedom promises insight into fundamental physics as well as the development of innovative quantum technologies. As mechanical resonators are massive and macroscopic, they can probe quantum theories at large scales \cite{Arndt2014,Viennot2018,Schoelkopf2018}, while the ability of mechanical motion to couple to a variety of quantum systems has inspired numerous mechanics-based transduction schemes \cite{Andrews2014,Maletinsky2014,Lukin2015,Nakamura2017,Groblacher2018,Safavi-Naeini2018,Awschalom2019}. Additionally, mechanical elements are compact compared to their electromagnetic counterparts, enabling the on-chip fabrication of many wavelength microwave structures such as high-performance filters and multimode resonators \cite{Morgan2007,Aref2016,Renninger2018}. High-fidelity control over the large number of modes achievable in acoustic platforms would be a powerful resource for quantum information processing \cite{Pechal2018}.

The field of circuit quantum electrodynamics (cQED) has provided both guidance and tools for achieving quantum control over mechanical excitations. In cQED, the state of a photonic mode is measured and manipulated using superconducting qubits. These qubits can also interact with mechanical systems using piezoelectric materials. Two seminal works leveraged this fact to couple a qubit to a dilatational resonator \cite{Cleland2010} and to propagating surface acoustic waves (SAWs) \cite{Delsing2014}. Both surface and bulk acoustic waves can be confined to form high-overtone resonators \cite{Leek2016,Rakich2018}, leading to demonstrations of qubit-phonon coupling in multimode cavities \cite{Leek2017,Schoelkopf2017,Lehnert2018,Astafiev2018,Sillanpaa2018}. Most recently, a pair of experiments used resonant interactions to create number and superposition states of an acoustic cavity mode, thereby demonstrating basic quantum control of acoustic phonons \cite{Cleland2018,Schoelkopf2018}. Following the example of cQED, achieving strong dispersive interactions in acoustic systems would lead to improved quantum control through quantum nondemolition phonon measurement \cite{Schuster2007,Johnson2010} and qubit mediated phonon-phonon interactions \cite{Wang2016,Schuster2017}. Realizing these techniques in acoustic systems would enable the exploration of a multimode analogy of cQED \cite{Heeres2017}. 


But coupling a single qubit with uniform strength to multiple modes of an acoustic cavity reduces the number of coherent interactions achievable with a given mode. Consider that for any qubit frequency inside the cavity bandwidth, there exists some nearest mode $k$ with detuning $\Delta_k$ less than half the cavity's free spectral range $f_s$. To be in the dispersive limit for all modes, the qubit must have coupling $g\ll\Delta_k< f_s/2$. This limited coupling then bounds the number of operations possible within the qubit's coherence time $(2\pi \gamma)^{-1}$ at approximately $g^2/(\Delta_k \gamma)$. The number of interactions achieved for mode $k$ is further reduced by a large factor $\approx|n-k| f_s/\Delta_k$ for the $n{\mathrm{th}}$ cavity mode. Eschewing the dispersive limit by choosing $g\approx f_s$ yields strong, resonant interactions between the qubit and multiple cavity modes. The resulting hybrid modes are composed predominantly of linear cavity modes, effectively diluting the qubit's nonlinearity and thereby increasing the time required for coherent operations \cite{Lehnert2018}. This reduction in coherent operations can be overcome by engineering a frequency-structured interaction such that modes far from the qubit frequency couple with rates exceeding $f_s$, while the coupling to nearby modes is suppressed, preserving the dispersive limit.

Indeed, acoustic platforms excel at realizing such strongly frequency-dependent couplings. In SAW devices, an interdigitated transducer (IDT) converts between electrical and acoustic signals with a frequency dependence determined by the Fourier transform of the IDT geometry \cite{Johansson2014,Aref2016}. A desired frequency response can be engineered by computing its inverse Fourier transform and shaping the IDT accordingly. Moreover, the slow speed of sound ($v=2880$\,m/s on GaAs) implies that megahertz frequency resolution can be realized with millimeter geometries. 

\begin{figure}[bt!]
	\centering
	\includegraphics[scale=1]{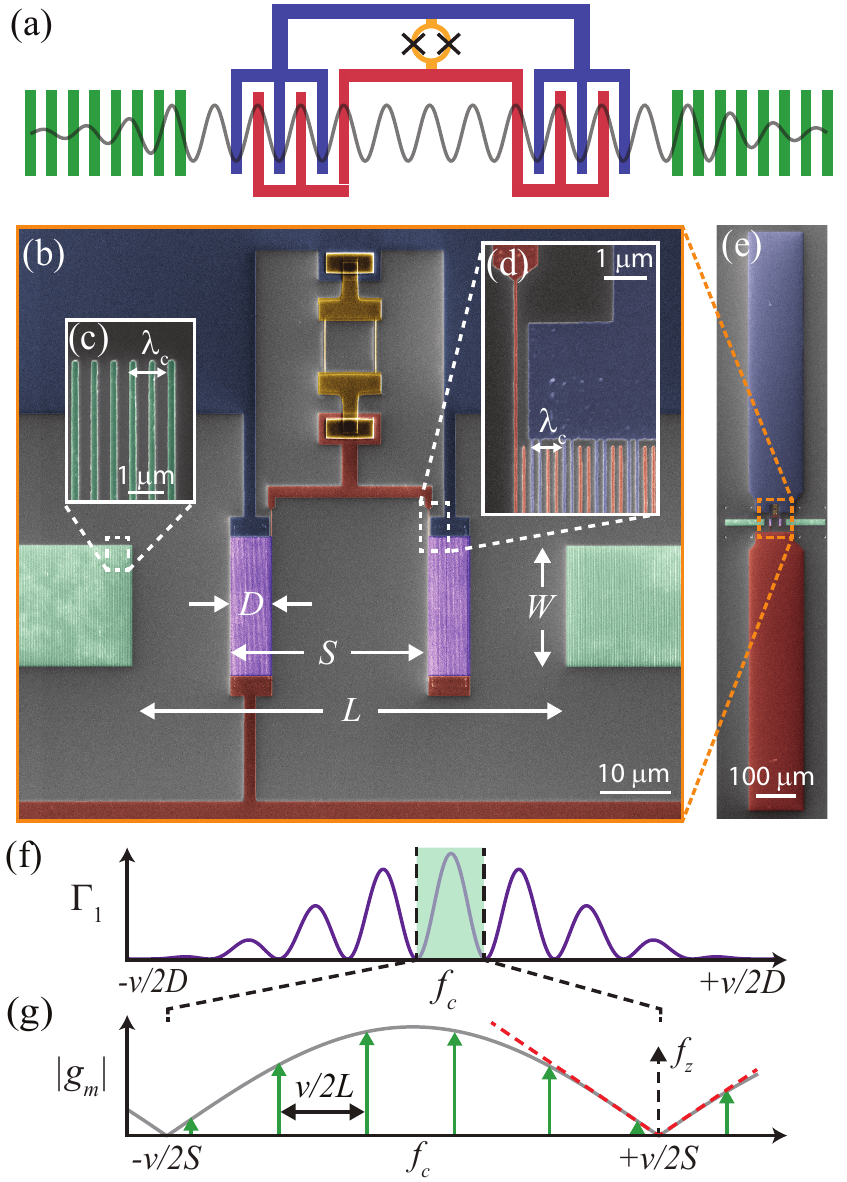}
	\caption{{\bf Double-slit qubit concept and device.} (a) Schematic of the device. A transmon qubit resembling a double-slit interferometer is formed by shunting two halves of an interdigitated transducers (IDT, red/blue) with a pair of Josephson junctions (yellow). Two distributed Bragg arrays (green) confine phonons in a narrow frequency band to create acoustic resonances (gray curve). (b)-(e) False color SEM micrographs of a representative device. (b) The qubit IDT (purple) is shunted by Josephson junctions and located within an acoustic cavity of width $W=16\,\upmu$m. The imaged device has half the cavity length of the measured device. Insets show enlarged images (c) of the Bragg mirror and (d) the IDT, whose innermost fingers are extended to shunt the junctions. (e) Antenna paddles couple the qubit to a copper waveguide cavity at 5.9\,GHz for readout and control. (f) When the qubit is tuned outside the mirror bandwidth (unshaded), the IDT launches phonons with a frequency dependence determined by the Fourier transform of the IDT geometry, creating fringes in the qubit loss rate $\Gamma_1$. (g) Inside the mirror band, the IDT modifies the qubit coupling strength $g_m$ to the evenly spaced cavity modes. By tuning the qubit frequency to a zero in the coupling at $f_z$, coupling rates exceeding the mode spacing can be achieved with dispersive operation, provided the slope near $f_z$ (red dotted line) is  much smaller than one.}
	\label{fig:schematic}
\end{figure}

In this article, we engineer a frequency-dependent coupling between a transmon qubit and a multimode SAW cavity to realize $g\sim f_s$ together with dispersive operation. The qubit couples to phonons through an IDT that is bisected to create a pair of interaction regions separated by a long travel time, $\tau \approx 9$\,ns [Fig.\,\ref{fig:schematic}(a)]. In close analogy to double-slit interference, the many-wavelength separation between interaction regions creates sharp fringes in the frequency dependence of the qubit-phonon interaction strength \cite{Johansson2014,Nori2018}. We observe the designed frequency dependence as a high-contrast modulation of both the coherent exchange rate between the qubit and cavity modes and the qubit spontaneous emission rate into unconfined phonons. This frequency dependence greatly reduces the coupling to certain modes to create frequency windows for dispersive operation. We tune the qubit transition to such a window and observe the single-phonon Stark shift from three strongly coupled modes of the cavity by populating these modes while measuring the qubit spectrum. For two of these modes, we enter the strong dispersive regime where the single-phonon Stark shift exceeds the qubit and acoustic linewidths, demonstrating that spatially extended coupling can be leveraged to take full advantage of multimode acoustic systems.

The device we study comprises a tunable transmon qubit on a piezoelectric GaAs surface with two IDT halves embedded in a multimode SAW cavity [Fig.\,\ref{fig:schematic}(a)]. The cavity is formed between two Bragg mirrors made of aluminum strips that reflect surface waves over a 100-\,MHz bandwidth to form a phononic Fabry-Perot cavity [Fig.\,\ref{fig:schematic}(b) and \ref{fig:schematic}(c)]. The effective cavity length extends beyond the mirror separation $L=125$\,\micron by 20\,\micron from acoustic penetration into the mirror array to create a mode spacing $f_s\approx 10 $\,MHz. The mirrors and IDT were designed with periodicity $\lambda_c=675$\,nm, which corresponds to a center frequency near $f_c\approx4.25$\,GHz. The IDT halves, each 8 periods long, are mirror images of each other reflected across the center of the cavity and separated by $S=26\,\upmu$m. The mechanical loading effects on the resonator from the IDT are minimized by using thin metal (30\,nm of aluminum) and a split electrode design [Fig.\,\ref{fig:schematic}(d)] \cite{Morgan2007}. Qubit readout and control are enabled by attaching antenna paddles [Fig.\,\ref{fig:schematic}(e)] that strongly couple the qubit to a copper waveguide cavity at 5.9\,GHz (see Appendix \ref{sup:readout}).

An IDT split in half achieves a mode-selective coupling by creating a frequency profile $A(f)$ analogous to the spatial profile of double-slit diffraction. The IDT is split into two regions of length $D$ separated by distance $S$. The Fourier transform about the symmetry point between these two regions is real and the product of two factors: a slow sinc envelope centered on $f_c$ with period $v/D$ and a fast sinusoidal modulation with period $1/\tau=v/S$:
\begin{equation*}
A(f)= \text{sinc}\left[ \pi (f-f_c) D / v\right] \text{sin}( \pi f \tau).
\end{equation*}
Outside the mirror bandwidth, the qubit loses energy to propagating phonons at a rate $\Gamma_1 \propto A(f)^2$ [Fig.\,\ref{fig:schematic}(f)] \cite{Johansson2014,Lehnert2018}. Within the mirror band, the qubit exchanges excitations with confined acoustic modes [Fig.\,\ref{fig:schematic}(g)] described by the multimode Jaynes-Cummings Hamiltonian
\begin{equation}
    H/h=\frac{1}{2} f_q \sigma_z+\sum_m f_m a_m^\dagger a_m +g_m (a_m^\dagger \sigma^-+a_m\sigma^+),
    \label{eq:H}
\end{equation}
 where the qubit is described by Pauli matrices and transition frequency $f_q$, the cavity modes are described by annihilation (creation) operators $a_m$ ($a_m^\dagger$) and frequencies $f_m$, and the qubit and cavity couple with strength $g_m$. If the IDT is symmetric about the cavity center, then $g_m$ has the form $g_m = g_0 A(f_m) \approx g_0 \sin\left(\pi f_m \tau \right)$ where $g_0$ is the maximal qubit-cavity coupling strength and the slowly varying sinc is approximated as unity. With the designed separation between IDT halves, the coupling varies with a periodicity approximately equal to the mirror bandwidth ($1/\tau\approx100$ \,MHz), ensuring at least one cavity mode achieves coupling near $g_0$.

Dispersive operation can be achieved regardless of mode density or maximal coupling strength by designing $A(f)$ to cross zero with a sufficiently shallow slope. Consider tuning the qubit to a frequency $f_z$ such that $A(f_z)=0$. A cavity mode with small detuning $\Delta_z$ from the qubit will couple with rate $g_z$ that is bounded above by this detuning multiplied by the slope of the coupling strength near $f_z$, i.e., $|g_z|<g_0 |A'(f_z) \Delta_z|$ [Fig.\, \ref{fig:schematic}(g)]. Thus, the magnitude of $A'(f_z)$ constrains $g_z/\Delta_z$ and thereby sets a lower limit on how dispersive qubit-cavity interactions can be. We engineer this slope, approximated for the split-IDT design as $g_0 A'(f_z)\approx \pi g_0\tau=0.14$, to be much smaller than one.

\begin{figure}[t]
	\centering
	\includegraphics[scale=1]{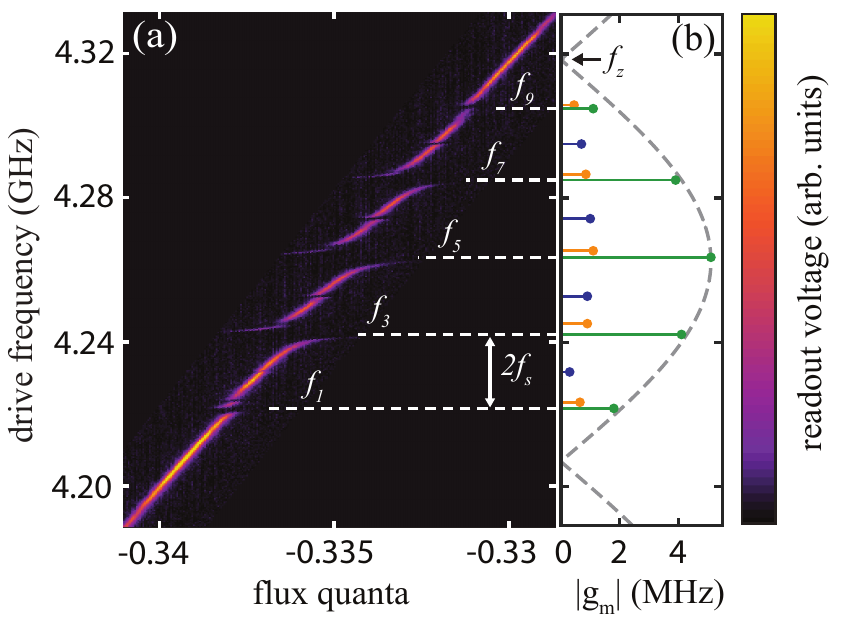}
	\caption{{\bf Avoided crossings spectroscopy} Tuning the qubit across the cavity modes measures the mode frequencies and qubit-cavity coupling strengths. (a) Within the mirror bandwidth, 9 avoided crossings appear with varying coupling strength and a nearly even mode spacing $f_s=10.6$\,MHz. The strong avoided crossings have a spurious mode on their high-frequency shoulder, which we attribute to a resonance with nonzero transverse mode number. (b) The coupling strengths to each mode are extracted from the measured crossings. The qubit couples more strongly to modes at the center of the mirror band with maximum strength $g_0=5.1$\,MHz. The IDT is symmetric about the cavity center, resulting in strong coupling to even modes (green) and weak coupling due to odd modes (blue). Coupling to transverse modes (orange) is a factor of 5 smaller. An inference of the frequency dependence of the qubit coupling strength (gray dotted line) made from measurements of the qubit spontaneous emission rate (see main text) agrees well with the measured coupling rates.}
	\label{fig:cavity}
\end{figure}

\begin{figure}[t]
	\centering
	\includegraphics[scale=1]{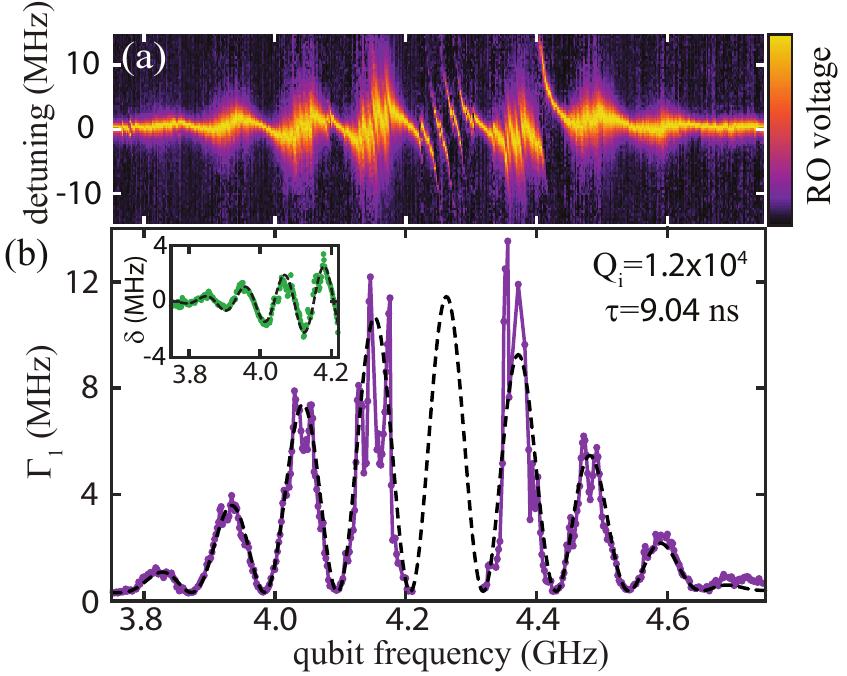}
	\caption{{\bf Interaction with propagating phonons} (a) Tuning the qubit across the $\sim1$-GHz IDT bandwidth probes its interaction with the continuum of propagating phonons outside the mirror band. Subtracting the expected flux tuning without acoustic interactions makes clear an oscillatory feature in both linewidth and frequency as seen in the measured readout (RO) voltage. Zigzag features bordering the central coherent avoided crossings are well described as resonant interactions with lossy acoustic modes. A strongly coupled system is present at 4.41\,GHz that is not a SAW-cavity mode. (b) The decay rate $\Gamma_1$ is found by measuring the qubit $T_1$ decay time. The measured energy loss rate varies by more than a factor of 25 within a 55-MHz span. A model closely fits the measurement except in the regions with the largest $\Gamma_1$, where increased uncertainty from the short decay times compounds with effects from the semireflective mirrors that are not included in the model. The measured frequency-dependent Lamb shift $\delta$ agrees well with the calculations from the IDT model (inset).}
	\label{fig:free}
\end{figure}

To confirm the designed frequency structure in the device, we measure the qubit spectrum as an applied magnetic flux tunes its frequency. We begin by tuning the qubit across the mirror bandwidth to investigate the frequency region where phonons are confined. We observe pronounced avoided crossings in the qubit spectrum where the qubit coherently exchanges energy with cavity phonons [Fig.\,\ref{fig:cavity}(a)]. The extracted coupling rates [Fig.\,\ref{fig:cavity}(b), see Appendix \ref{sup:crossings}] vary between the modes, with several strongly coupled modes in close spectral proximity to crossing-free regions wider than $f_s$ at both edges of the mirror band. Three main effects explain the observed behavior. First, the split-IDT modulates the coupling proportional to $A(f)$, coupling the qubit strongly to modes near 4.25\,GHz with $g_0=5.1$\,MHz while decoupling it from modes roughly $5f_s$ above or below. Second, neighboring cavity modes couple to the qubit with alternating strength because the qubit IDT is approximately symmetric about the cavity center, strongly (weakly) coupling the qubit to modes with even (odd) spatial symmetry. Lastly, resonant exchange between the qubit and cavity modes at the edge of the mirror bandwidth is unresolved as the coupling rate is much less than the loss rate of these weakly confined modes. 

To study $A(f)$ outside the mirror band, we tune the qubit over a 1-GHz span and examine the influence of propagating phonons on the qubit linewidth and transition frequency. In contrast to the discrete cavity modes, propagating modes form a continuum, enabling a dense sampling of $A(f)$ over a broad frequency range and affording a clear picture of how effectively the split IDT tailored the qubit-phonon interaction. In the measured qubit spectra [Fig.\,\ref{fig:free}(a)], the features arising from acoustic interactions are emphasized by subtracting the flux dependence expected from an acoustically uncoupled qubit (see Appendix \ref{sup:flux}). At frequencies detuned from the central avoided crossings, the qubit linewidth oscillates with a period of 110\,MHz that is consistent with the expected delay time and an amplitude that decays as the qubit tunes out of the IDT bandwidth. Additionally, the qubit frequency deviates from the uncoupled flux dependence with a similarly enveloped oscillation with matching 110-MHz periodicity. Both of these effects can be understood by modeling the qubit's emission of phonons from the IDT as a frequency-dependent resistance, which must be accompanied by a frequency-dependent reactance from Kramers-Kronig relations \cite{Morgan2007,Aref2016}. We observe this reactance as a modulation of the qubit frequency compared to its uncoupled flux tuning, an effect describable as a phononic Lamb shift \cite{Wang2015,Johansson2014}. 

We determine the qubit energy decay rate with increased precision by measuring qubit excited state lifetime ($T_1$) in the time domain. With the qubit far detuned from the acoustic cavity modes, we observe $\Gamma_1= (2\pi T_1)^{-1}$ oscillating in frequency with large amplitude; the loss increases by a factor of 25 above its minimal value within a $55$-MHz span [Fig.\,\ref{fig:free}(b)]. A simple model that combines a prediction for the phonon emission rate from the IDT and a constant internal quality factor $Q_i$ closely fits the measured qubit loss rate, giving $Q_i=1.2\times10^4$ and $\tau=9.04$\,ns (see Appendix\,\ref{sup:T1}). The nulls in $\Gamma_1$ arise from destructive interference between the two IDT halves, an effect with close parallels to an atom interfering with its mirror image \cite{Eschner2001,Wilson2015}. As the depth of these nulls is approximately uniform across the IDT bandwidth, phonon loss from imperfect destructive interference is less than 75\,kHz. Additionally, the extracted IDT parameters from the qubit loss rate can be used to calculate the frequency-dependent phononic Lamb shift, showing agreement with the measured qubit frequency [inset of Fig.\,\ref{fig:free}(b)].

Our measurement of the qubit interaction with propagating modes also provides an independent inference of the interaction strength between the qubit and cavity modes. The best-fit model from Fig.\,\ref{fig:free}(b) determines $A(f)$ using propagating modes and can be extended to frequencies inside the mirror band, where it closely follows the measured coupling rates [Fig.\,\ref{fig:cavity}(b)]. 

\begin{figure}[bt]
	\centering
	\includegraphics[scale=1]{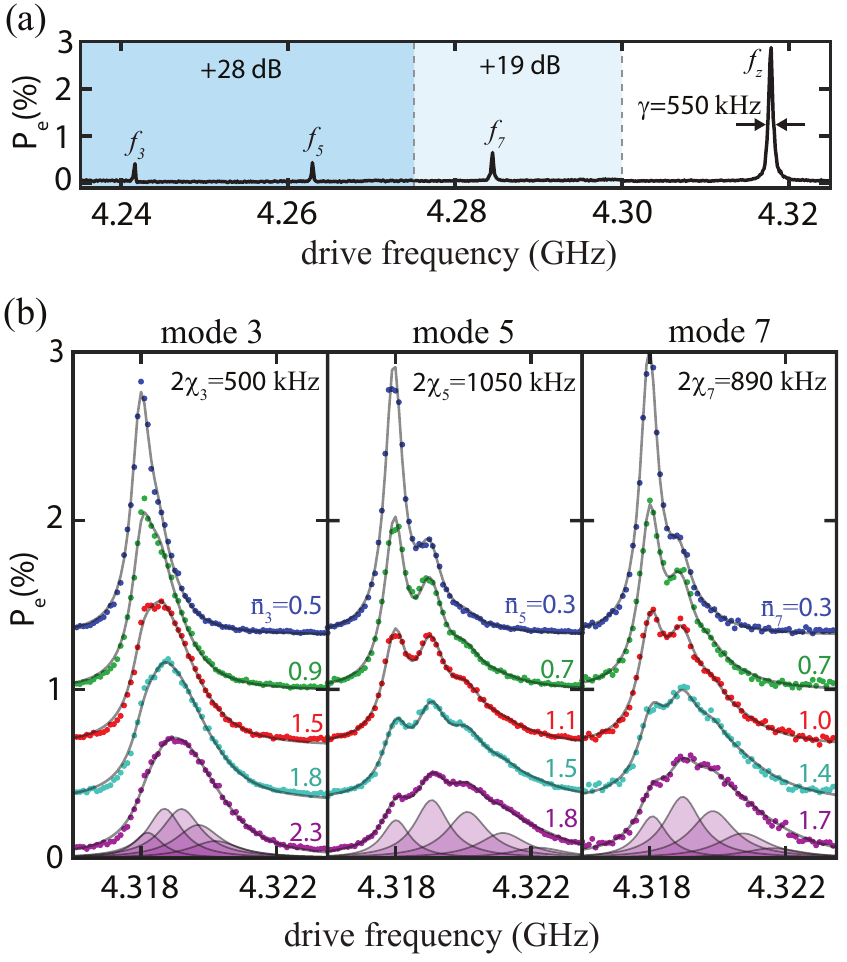}
	\caption{{\bf Phonon number splitting} The qubit is tuned to $f_z= 4.318$\,GHz where no resonant interactions occur and the first two transmon transitions straddle the acoustic modes. (a) To excite and measure the acoustic modes, the drive power is increased relative to the power used near the qubit transition (shaded, +19 or +28 dB). The interactions with all acoustic modes are dispersive, with $\Delta_{3}/g_{3}=18$, $\Delta_{5}/g_{5}=11$, and $\Delta_{7}/g_{7}=8.5$. (b) While driving the strongly coupled acoustic modes (3, 5, and 7), we observe the qubit transition broaden and shift up in frequency with increasing phonon population. The spectra are fit to a model assuming a coherent state with $\overline{n}_m$ average phonons to determine $\chi_m$. For modes 5 and 7, the measured single-phonon Stark shift exceeds both the qubit and acoustic linewidths. Traces are vertically offset to aid clarity. For the trace in each mode with the highest phonon occupancy, the contributions to the best fit from each phonon Fock state are shown (shaded purple).} 
	\label{fig:numb}
\end{figure}

Having characterized the qubit interaction with both confined and propagating phonons, we turn to resolving the qubit's Stark shift from individual cavity phonons. This resolution requires dispersive operation with all modes, i.e. $g_m \ll |\Delta_m|$ for all modes $m$, where $\Delta_m=f_q-f_m$, as well as a Stark shift that exceeds both the qubit and acoustic loss rates. Tuning the qubit to $f_z=4.318$\,GHz realizes dispersive operation; the least-dispersive interaction is with mode 7 where $\Delta_7/g_7=8.5\gg 1$. In this multimode dispersive regime, the interaction term in the Hamiltonian [Eq. \ref{eq:H}] becomes 
\begin{equation*}
    H_I/h=\sum_m \chi_m a_m^\dagger a_m \sigma_z,
\end{equation*}
where individual phonons shift the transmon frequency by $2\chi_m$. An accurate calculation of $\chi_m$ must include the higher levels of the transmon, and is well approximated as
\begin{equation}
\chi_m=g_m^2\left(\frac{1}{\Delta_m}-\frac{1}{\Delta_m+\alpha} \right),
\label{eq:chi}
\end{equation} 
where $\alpha=-190$\,MHz is the transmon's anharmonicity. With the qubit at $f_z$, its transition frequency is above the acoustic modes while the $|e\rangle\rightarrow|f\rangle$ transition is below, such that $\Delta_m>0$ and $\Delta_m+\alpha<0$ for all modes $m$. With this level ordering, the two terms in Eq.\,(\ref{eq:chi}) add constructively to create large and positive Stark shifts \cite{Koch2007}.

To populate a target cavity mode with phonons, we drive the qubit at a frequency far detuned from its own transition but resonant with the cavity mode \cite{Schoelkopf2018}. In Fig.\,\ref{fig:numb}(a), spectroscopy shows the qubit transition at 4.318\,GHz and, with much higher drive power, three acoustic resonances at lower frequencies. The measured qubit linewidth $\gamma=550$\,kHz is only marginally larger than the sum of contributions from $Q_i$, intrinsic dephasing, and expected power broadening (see Appendix \ref{sup:numb}). The acoustic linewidths are measured to be $\kappa_m\approx 250$\,kHz for all three modes, only slightly larger than the expected 200\,kHz of diffraction loss from the flat-flat mirror design of the cavity \cite{Leek2016,Lehnert2018,Aref2016}. 

We measure the single-phonon Stark shift of the three strongly coupled modes by varying the population in these modes and measuring the qubit spectrum. A 3-\micro s drive pulse at $f_m$ creates a coherent state in mode $m$ with $\overline{n}_m$ average phonons \cite{Schuster2007}. The resulting Stark-driven qubit spectrum, measured with a spectroscopy pulse concurrent with the acoustic drive, consists of a sum of Lorentzians that each correspond to a phonon number state in the cavity [Fig.\,\ref{fig:numb}(b)]. These Lorentzians are spaced by $2\chi_m$ and broaden with higher phonon number in proportion to $\kappa_m$. Sweeping the drive power at one of the three modes, the measured qubit spectrum broadens and shifts up in frequency. Crucially, several resolved peaks appear for modes 5 and 7 arising from a distribution of phonon Fock states in the cavity. To model the measurement, we assume the cavity occupation is Poissonian distributed and fit the average phonon number in each trace. We find good agreement between the model and measurement for acoustic linewidths $\kappa_{3,5,7}=200,250,275$\,kHz and single-phonon Stark shifts $2\chi_{3,5,7}=500,1050, 890$\,kHz (see Appendix\,\ref{sup:numb}). As the single-phonon Stark shifts for modes 5 and 7 exceed both the qubit and acoustic linewidths, we confirm that the device enters the strong dispersive regime for two acoustic modes.

Resolving phonon Fock states in a multimode cavity through spatial engineering suggests multiple future directions. For the measured device, the dominant source of phonon loss was likely diffraction and could be eliminated by using curved reflectors to form a stable cavity \cite{Awschalom2019}. Combining improved phonon lifetimes with the demonstrated coupling strengths would enable quantum nondemolition phonon detection and qubit-mediated interactions between phonon modes. Furthermore, the number of modes accessible to the qubit can be increased simply by elongating the cavity, highlighting the promise of SAW systems for multimode quantum information processing \cite{Schuster2017,Schoelkopf2018}. More generally, the engineering of time-delayed self-interactions not only enables a wide range of frequency structures but can also give rise to non-Markovian dynamics \cite{Johansson2017,Delsing2018}, suggesting delay may prove a valuable resource for quantum information processing \cite{Lukin2017}.

See related work Ref. \cite{Amir2019}.

	\section*{Acknowledgements} 
	
We thank Xizheng Ma for insightful discussions as well as Daniel Palken and Maxime Malnou for providing the quantum-limited amplifier. This work was supported by NSF Grant No. 1734006.

\appendix

\section{Qubit readout}
\label{sup:readout}

The qubit state is measured through its dispersive interaction with a 5.9-GHz copper waveguide cavity. The qubit has a large electric dipole moment, coupling it to the readout cavity with strength $g_c=115$\,MHz. Different readout techniques were used to probe the qubit state depending on the measurement details.

We used bright-state readout \cite{Schoelkopf2010} to measure the qubit decay rate $\Gamma_1$ as a function of frequency [Fig.\,\ref{fig:free}(b)]. This type of readout is well suited for measuring fast decays as the cavity can persist in the bright state for a time that exceeds the natural qubit lifetime.

For qubit spectroscopy, we used single quadrature dispersive readout backed by a flux-pumped Josephson parametric amplifier. To measure the Stark-driven qubit spectra, we used a pulsed readout scheme that minimized qubit dephasing from readout phonons [Fig.\,\ref{fig:numb}]. Continuous readout was used for qubit spectroscopy as a function of flux [Figs.\,\ref{fig:cavity}(a) and \ref{fig:free}(a)]. In the broad qubit spectroscopy, we compensate for the varying excited state contrast resulting from frequency-dependent qubit loss by adjusting the qubit drive power. This power level is independently determined from the measured $T_1$ times.

\section{Qubit flux dependence}
\label{sup:flux}

\begin{figure}[tb]
	\centering
	\includegraphics[scale=1]{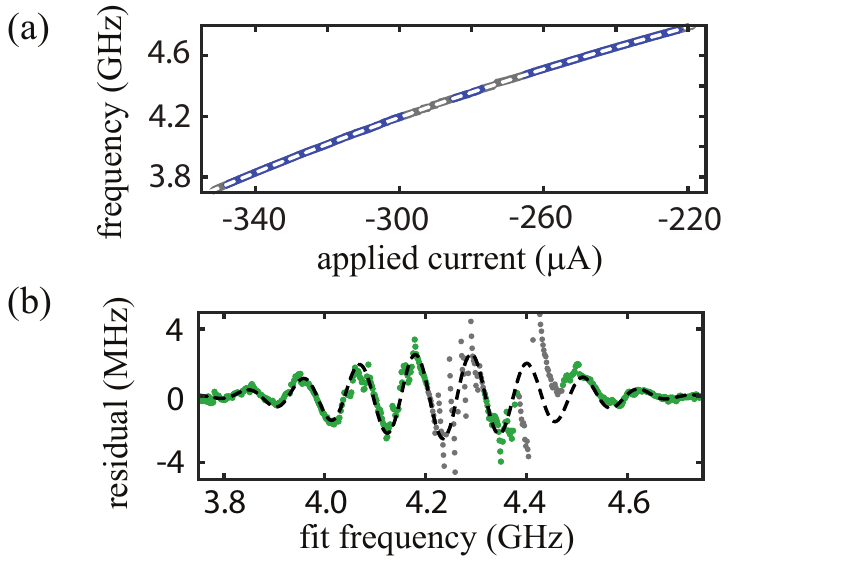}
	\caption{{\bf Qubit flux dependence} (a) The qubit frequency is measured and fit (white line) over a large range of applied flux. Regions near avoided crossings are ignored in the fit (gray). (b) The residual between the measured transitions and the best-fit uncoupled flux dependence matches the expected acoustic Lamb shift (black).}
	\label{fig:flux}
\end{figure}

The qubit transition frequency is tuned using an off-chip coil to thread magnetic flux through the 50-$\upmu\text{m}^2$ loop formed by the two Josephson junctions. Omitting acoustic interactions, we model the qubit frequency $f_q$ as a function of coil current $I$ as
\begin{equation*}
f_q(I)=f_0\left[a^2+(1-a^2) \cos\left(\pi\frac{I-I_0}{I_c}\right)^2\right]^{1/4}, 
\end{equation*}
where $f_0$ is the zero-field qubit frequency, $I_c$ is the coil current required to thread a half flux quantum through the qubit loop, $I_0$ is the current offset required to offset ambient fields, and $a$ is the normalized difference between the junction critical currents. From fitting the measured qubit frequency [Fig.\,\ref{fig:flux}(a)], we find $f_0=5.718$\,GHz, $I_c=1.168$\,mA, $I_0=79.2$\,\micro A, and $a=0.14$. 

The qubit flux dependence is weakly modified by its interaction with the continuum of propagating phonon modes. We model this phononic Lamb shift $\delta$ as
\begin{equation*}
\delta(f_q)=\frac{\Gamma_0}{4}\text{sinc}\left(\pi N_q \frac{f_q-f_c}{f_c}\right)^2 \text{sin}(\pi f_q \tau)
\end{equation*}
where $\Gamma_0$ is the maximal loss rate to phonons, $f_c$ is the center frequency of the IDT, $N_q=8$ is the number of finger periods in each IDT, and $\tau$ is the intra-IDT delay \cite{Johansson2014}. The measured phonon loss rate (see Appendix \ref{sup:T1}) independently determines $\Gamma_0$, $f_c$, and $\tau$, allowing the Lamb shift to be calculated with no free parameters. This calculated Lamb shift closely matches the residual from the flux fit [inset of Fig.\,\ref{fig:free}(b) and Fig.\,\ref{fig:flux}(b)] except near avoided crossings. 
	
\section{Acoustic cavity characterization}
\label{sup:crossings}

\begin{figure}[b]
	\centering
	\includegraphics[scale=1]{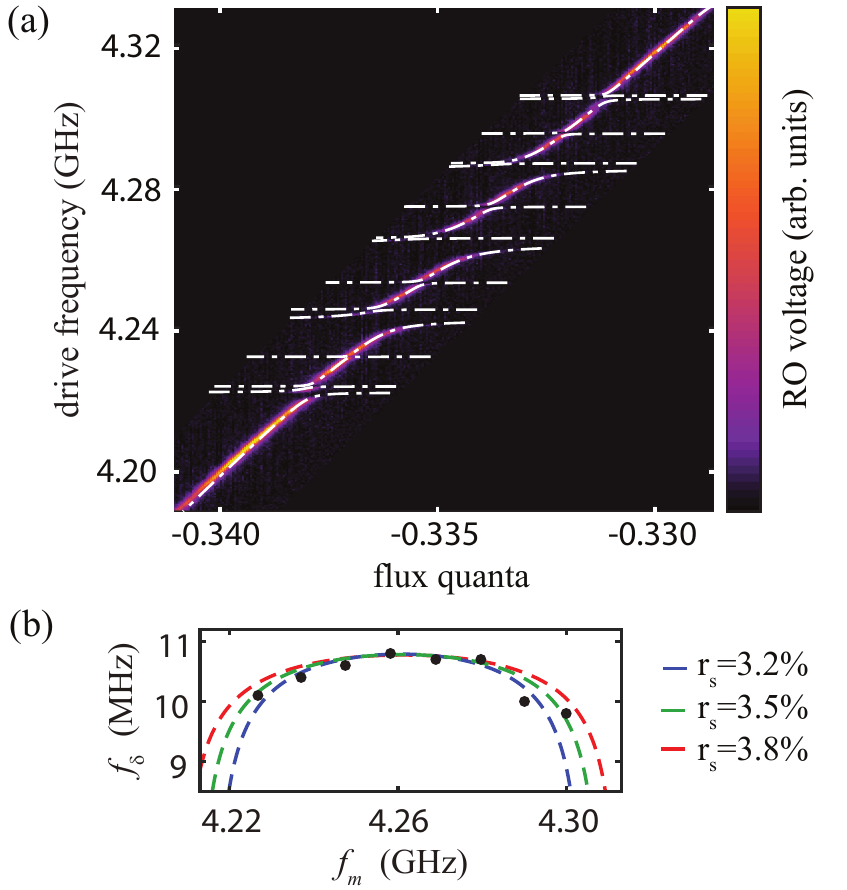}
	\caption{{\bf Avoided crossings analysis.} (a) The measured qubit-cavity avoided crossings are fit by a multimode interaction model (white) [Fig.\,\ref{fig:cavity}]. (b) The spacings between the modes $f_\delta$ are well matched by a mirror model with a single-element reflectivity of $r_s=3.5\%$ (green).}
	\label{fig:crossings}
\end{figure}

Extracting the coupling strengths from the closely spaced avoided crossings requires a multimode formalism. The eigenmodes of the system are found by diagonalizing the interaction Hamiltonian,
\begin{equation*}
 H/h = \begin{pmatrix} f_1 & & & g_1\\ 
 & f_2 & & g_2\\ 
 & & \ddots & \vdots \\ 
 g_1 & g_2 & \cdots & f_{\mathrm{q}}
 \end{pmatrix} ,
\label{hamiltonian}
\end{equation*}
including 9 purely longitudinal modes and 5 transverse modes. The eigenvalues of the matrices as a function of flux are fit to the measured avoided crossing spectrum [Fig.\,\ref{fig:crossings}(a)]. 

The general properties of the mirrors can be inferred from the precise measurement of the mode spacings. Near the center of the mirror bandwidth, the modes are spaced by $f_s=10.6$\,MHz, but they become more closely spaced near the edge of the mirror bandwidth due to deeper phonon propagation into the mirror stack [Fig.\,\ref{fig:crossings}(b)]. We find a simple mirror model matches the measurements with a single-element reflectivity $r_s=3.5\%$, which corresponds to a mirror bandwidth of 100\,MHz.

\section{Phonon emission rate}
\label{sup:T1}

The qubit lifetime is measured over a wide frequency range to directly probe the qubit spontaneous emission rate into unconfined phonons. The qubit loss rate $\Gamma_1$ as a function of qubit frequency $f_q$ is modeled by
\begin{equation*}
\Gamma_1(f_q)=\frac{f_q}{Q_i}+\frac{\Gamma_0}{2}\text{sinc}\left(\pi N_q \frac{f_q-f_c}{f_c}\right)^2[1-\cos(2\pi f_q \tau)],
\end{equation*}
where $Q_i$ is the qubit internal quality factor, $\Gamma_0$ is the maximal loss rate to phonons, $f_c$ is the center frequency of the IDT, $N_q=8$ is the number of finger periods in each IDT, and $\tau$ is the intra-IDT delay time. We find $Q_i=1.2\times10^4$, $\Gamma_0=11$\,MHz, $f_c=4.24$\,GHz, and $\tau=9.04$\,ns. The best fit $\Gamma_0$ is close to the expected value of 12.5\,MHz calculated using room temperature GaAs properties \cite{Delsing2014}. 

\begin{figure}[tb]
	\centering
	\includegraphics[scale=1]{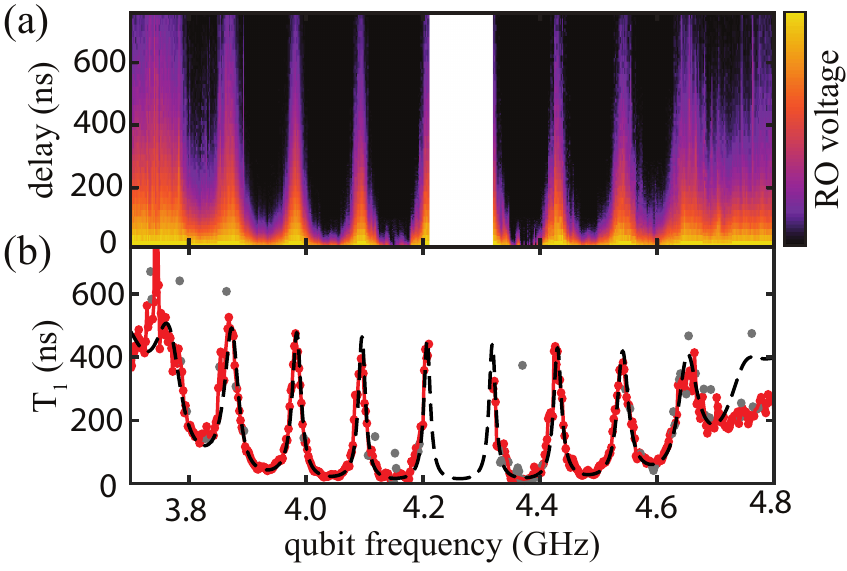}
	\caption{{\bf Qubit spontaneous emission.} (a) Each measured qubit $T_1$ decay is shown normalized by its maximum value. (b) The extracted $T_1$ times match the model for the qubit-phonon emission rate. A small fraction of decays displayed nonexponential features with timescales inconsistent with the intra-IDT delay time (gray).}
	\label{fig:T1vsF}
\end{figure}

The qubit studied constitutes a giant atom where the intra-IDT delay time approaches the phonon-limited qubit lifetime. Deep in this regime, the qubit fully decays before a phonon can travel between the IDT halves, leading to a host of effects such as nonexponential decay. The transition to this regime occurs when the product $\pi \tau \Gamma_0$ reaches 1 \cite{Johansson2014,Delsing2018}. For this device, $\pi \tau \Gamma_0 \approx0.3$. However, evidence of non-Markovian physics was obscured by the presence of mirrors and the short timescale (9\,ns) associated with the nonexponential decays. A small fraction of the measured time traces display nonexponential features but with timescales far exceeding the intra-IDT delay time. These decays are excluded from the reported qubit energy decay rates [Fig\,\ref{fig:T1vsF}].

\section{Number splitting analysis}
\label{sup:numb}

\begin{figure}[b!]
	\centering
	\includegraphics[scale=1]{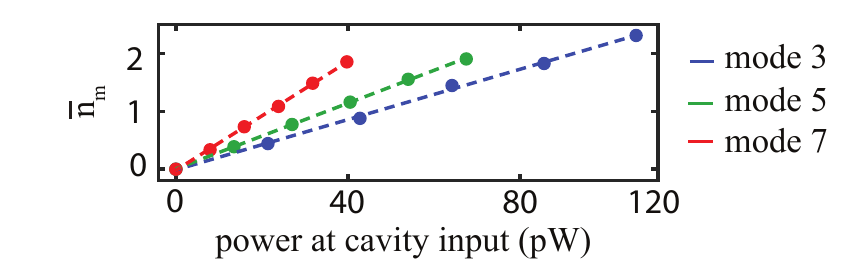}
	\caption{{\bf Phonon number power dependence} The average phonon number extracted from the measurements depends linearly on the power applied to each mode.}
	\label{fig:NumbMeta}
\end{figure}

The measured Stark-driven spectra are fit to a sum of unit-area Lorentzians with weights assumed to be Poissonian distributed in the number basis with mean phonon number $ \overline{n}$,
\begin{equation*}
 P_e(f, \overline{n})=C_0+C_1\sum_{n=0}^{n_{\text{max}}} P_n( \overline{n}) S(f, \overline{n},n),
\end{equation*}
where $n$ is the phonon number in mode $m$, $f$ is the spectroscopy frequency, $C_0$ is a constant offset, $C_1$ is an overall amplitude, and $n_{\text{max}}=6$ is a cutoff phonon number. The two factors in the sum are given by
\begin{align*}
 &P_n( \overline{n})=e^{- \overline{n}} \frac{ \overline{n}^n}{n!}\quad \quad \text{and}\\
 &S=\frac{1}{2\pi}\frac{\gamma+\kappa_m (n+ \overline{n})}{[f-(f_q-2\chi_m n)]^2+[\gamma+\kappa_m(n+ \overline{n})]^2/4},
\end{align*}
where $\gamma$ is the zero-phonon qubit linewidth, $f_q$ is the zero-phonon qubit frequency, $\kappa_m$ is the loss rate of mode $m$, and $2\chi_m$ is the single-phonon Stark shift from mode $m$. Fits of the average phonon number show a linear dependence on applied drive power for the three measured modes [Fig.\,\ref{fig:NumbMeta}]. The strong drive used to populate the acoustic modes also weakly excites the qubit, causing the trace offset $C_0$ to increase with $ \overline{n}$. Additionally, the bare qubit frequency pulls weakly up with off-resonant drive power at a rate of about 150\,kHz per phonon, an unexplained effect that is included in the fits.

The qubit coherence times at $f_z$ are measured to be $T_1=415$\,ns and $T_2^*=705$\,ns. The $T_2^*$ time is almost twice $T_1$, and we calculate an intrinsic dephasing rate of $(2\pi T_{\phi})^{-1}=30\,$kHz. The spectroscopic qubit linewidth was measured to be $\gamma=550$\,kHz at $f_z$. Together, frequency-independent energy loss (360\,kHz), intrinsic dephasing (30\,kHz), the effective Rabi rate from the drive tone (100\,kHz), and the finite duration of the drive pulse (50\,kHz) sum to a 540-kHz qubit linewidth. marginally smaller than the measured value.

Additionally, an unstable avoided crossing appeared intermittently between 4.312 and 4.322\,GHz with sub-MHz coupling rate, fluctuating with a several-hour timescale. We reject data when the defect was present by interleaving independent diagnostics with the Stark-driven spectra and removing defect-present data in postprocessing.

\end{document}